\documentclass[%
 reprint, superscriptaddress,
 amsmath,amssymb,
 aps,
 prl,
]{revtex4-1}

\usepackage{graphicx}
\usepackage{dcolumn}
\usepackage{bm}
\usepackage{soul}

\usepackage[usenames,dvipsnames]{xcolor}

\usepackage{hyperref}
\hypersetup{
  colorlinks,
  allcolors=NavyBlue,
  linktoc=all
}

\newcommand*{\unit}[1]{\ensuremath{\,\mathrm{#1}}}
\newcommand*{\s}[1]{\ensuremath{_\mathrm{#1}}}	
\newcommand*\C{\mathcal C}
\newcommand*\nba{\bar n\s{BA}}
\newcommand*\Geff{\Gamma\s{eff}}
\newcommand*\nth{\bar n\s{th}}
\newcommand*\nimp{\bar n\s{imp}}
\newcommand*\nadd{\bar n\s{add}}
\newcommand*\dagg{^{\dagger}}

\newcommand*\da{\delta\hat a}
\newcommand*\db{\delta\hat b}

\begin{document}

\preprint{APS/123-QED}

\title{Optical Backaction-Evading Measurement of a Mechanical Oscillator}

\author{Itay Shomroni}\thanks{These two authors contributed equally}
\author{Liu Qiu}\thanks{These two authors contributed equally}
\affiliation{Institute of Physics, \'Ecole Polytechnique F\'ed\'erale de Lausanne, Lausanne 1015, Switzerland}
\author{Daniel Malz}
\author{Andreas Nunnenkamp}
\affiliation{Cavendish Laboratory, University of Cambridge, Cambridge CB3 0HE, United Kingdom}
\author{Tobias J. Kippenberg}
\email{tobias.kippenberg@epfl.ch}
\affiliation{Institute of Physics, \'Ecole Polytechnique F\'ed\'erale de Lausanne, Lausanne 1015, Switzerland}

\date{September 4, 2018}

\begin{abstract}
Quantum mechanics imposes a limit on the precision of a continuous position measurement of a harmonic oscillator, as a result of quantum backaction arising from quantum fluctuations in the measurement field.
A variety of techniques to surpass this standard quantum limit have been proposed, such as variational measurements, stroboscopic quantum non-demolition and two-tone backaction-evading (BAE) measurements.
The latter proceed by monitoring only one of the two non-commuting quadratures of the motion.
This technique, originally proposed in the context of gravitational wave detection, has not been implemented using optical interferometers to date.
Here we demonstrate continuous two-tone backaction-evading measurement 
in the optical domain of a localized GHz frequency mechanical mode of a photonic crystal nanobeam cryogenically and optomechanically cooled in a $^3$He buffer gas cryostat close to the ground state.
Employing quantum-limited optical heterodyne detection, we explicitly show the transition from conventional to backaction-evading measurement.
We observe up to $0.67\unit{dB}$ (14\%) reduction of total measurement noise, thereby demonstrating the viability of BAE measurements for optical ultrasensitive measurements of motion and force in nanomechanical resonators.
\end{abstract}

\pacs{Valid PACS appear here}
\maketitle


In a continuous measurement of the position $\hat x$ of a harmonic oscillator, quantum backaction (QBA) of the measuring probe on the momentum $\hat p$ ultimately limits the attainable precision~\cite{caves1980,clerk2010}, restricting ultrasensitive measurements of force or motion.
For an interferometric position measurement, in which a mechanical oscillator is parametrically coupled to a cavity, the trade-off arising from measurement imprecision (i.e.~detector shot noise) and QBA force noise on the mechanical oscillator, dictates a minimum added noise equivalent to the oscillator's zero-point fluctuations, $x\s{zpf}=\sqrt{\hbar/2m\Omega_m}$, referred to as the Standard Quantum Limit (SQL), originally studied in the context of gravitational wave detection~\cite{pace1993,caves1980} (here $m$ is the mass, and $\Omega_m$ the angular frequency of the mechanical oscillator).
Recent advances in the field of cavity optomechanics~\cite{aspelmeyer2014} which utilizes nano- or micro-mechanical oscillator coupled to optical or superconducting microwave cavities, have allowed reaching the regime where the QBA arising from radiation pressure quantum fluctuations becomes relevant. 
In particular, imprecision noise far below that at the SQL has been obtained~\cite{teufel2009,anetsberger2010}, thus entering the QBA-dominated regime; QBA has been observed~\cite{murch2008,purdy2013,teufel2016}; and sensitivities approaching the SQL have been demonstrated~\cite{schreppler2014,wilson2015,teufel2016,rossi2018}.

Quantum non-demolition (QND) techniques, first proposed by Thorne, Braginsky \textit{et al.}~\cite{thorne1978,braginsky1980,caves1980b}, allow beating the SQL by minimizing or evading the effects of QBA.
One technique to surpass the SQL, applicable to measurements far from the mechanical resonance frequency $\Omega_m$, utilizes quantum correlations in the probe (due to ponderomotive squeezing~\cite{braginsky1967,purdy2013b,safavi-naeini2013,sudhir2017b}), known as `variational readout'~\cite{vyatchanin1995,kimble2001}. This technique has recently been demonstrated in a cryogenic micromechanical oscillator coupled to an optical cavity~\cite{kampel2017}, and in a room temperature nano-optomechanical system for quantum-enhanced force measurements~\cite{sudhir2017}.
Another possibility is utilizing squeezed light, a technique applied to gravitational wave detectors~\cite{jaekel1990,kimble2001,collaboration2011}.
More recent schemes include measurements of the collective motion in a hybrid system composed either of two mechanical oscillators (as demonstrated for an electromechanical system~\cite{woolley2013,ockeloen-korppi2016}), or a mechanical and a `negative mass' oscillator (demonstrated using an atomic ensemble~\cite{polzik2015,moller2017}).

\begin{figure}
\includegraphics[scale=1]{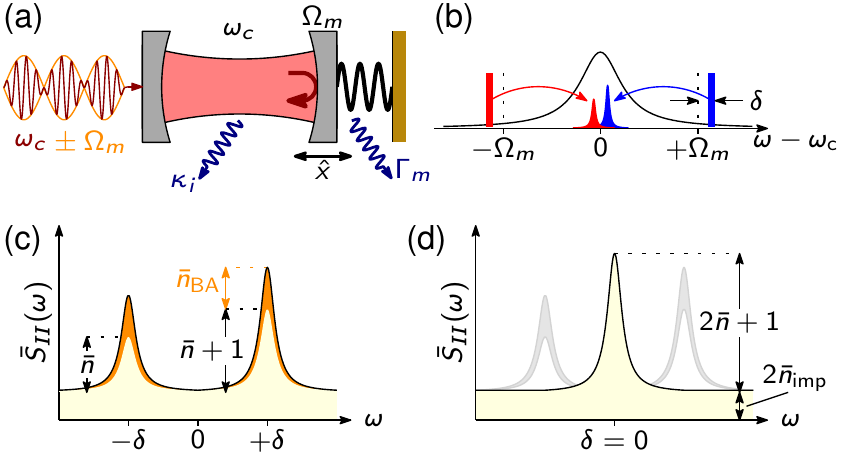}
\caption{\textbf{Backaction-evading measurement.}
(a)~Illustration of a cavity optomechanical system. 
Light in a cavity with optical resonance $\omega_c$ and full linewidth $\kappa$ (of which $\kappa_i$ intrinsic losses) is coupled to the position $\hat x$ of a mechanical oscillator that has frequency $\Omega_m$ and linewidth $\Gamma_m$. 
In a BAE measurement, the probe is amplitude-modulated at the mechanical frequency $\Omega_m$, coupling to the quadrature $\hat X$.
(b)~Frequency space configuration slightly detuned from BAE measurement, where the probe is modulated at $\Omega_m+\delta$.
(c)~Resulting power spectral density for an oscillator in a thermal state, showing the asymmetric Stokes and anti-Stokes scattered sidebands, plus the heating due to QBA.
(d)~When tuning to the BAE scheme $\delta=0$, the two sidebands coalesce and the QBA is cancelled, see Eq.~\eqref{eq:Shet}.
The remaining imprecision noise $\nimp$ can be arbitrarily reduced by increasing probe power.
}
\label{fig:psd}
\end{figure}
\begin{figure*}[th]
\includegraphics[scale=1]{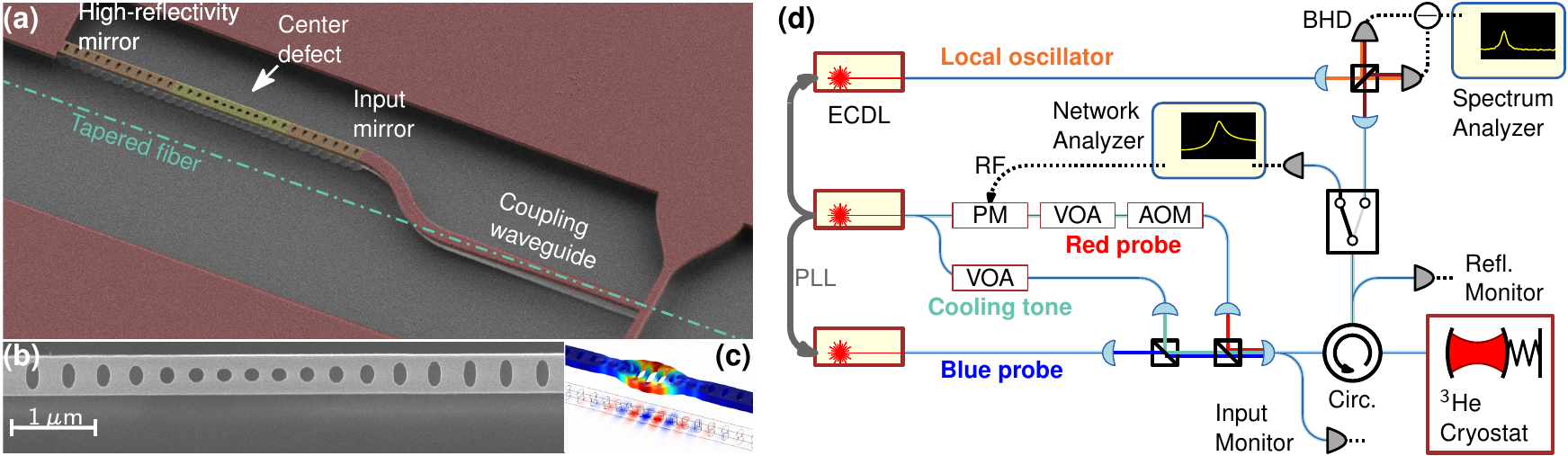}
\caption{
  \textbf{Optomechanical crystal and experimental setup.}
  (a)~False-color SEM image of the silicon optomechanical crystal cavity with a waveguide for laser input coupling.
  The path of the tapered fiber is indicated.
  (b)~SEM image detail of the cavity.
  (c)~Illustration of the mechanical breathing mode (top) and optical mode (bottom).
  (d)~Experimental setup. ECDL, external-cavity diode laser; PM, phase modulator; VOA, variable optical attenuator;
AOM, acousto-optical modulator; BHD, balanced heterodyne detector; PLL, phase-locked loop.
See Ref.~\onlinecite{qiu2018} for more details.
}
\label{fig:experimental_setup}
\end{figure*}

Another type of QND measurement, backaction-evading (BAE) measurements introduced by Thorne \textit{et al.}~\cite{thorne1978}, allow  avoiding QBA entirely by 
measuring only one of the two slowly-varying amplitude and phase quadratures $\hat X$ and $\hat Y$, defined by 
$\hat x(t) \equiv \sqrt{2}x\s{zpf}[\hat X(t)\cos\Omega_m t + \hat Y(t)\sin\Omega_m t]$, which constitute QND observables.
Unlike $\hat x$ and $\hat p$, the conjugate observables $\hat X$ and $\hat Y$
are \emph{decoupled} from each other during free dynamic evolution.
By exclusively measuring $\hat X$ (say), all QBA is diverted to $\hat Y$
and is completely absent from the measurement record.
By increasing coupling to the system (probe power), one can then
arbitrarily reduce the imprecision noise,
allowing in principle unlimited sensitivity in the measurement of one quadrature.
In a cavity optomechanical system
such backaction-evading (BAE) measurement is possible by amplitude-modulating a cavity-resonant probe at frequency $\Omega_m$ (Fig.~\ref{fig:psd}a)~\cite{braginsky1980,caves1980b,clerk2008},
equivalent to two-tone probing on the upper and lower mechanical sidebands of the cavity (Fig.~\ref{fig:psd}b)~\footnote{It has been pointed out~\cite{kampel2017} that single-quadrature measurement may also be realized using synodyne detection~\cite{buchmann2016}.}.
Two-tone BAE is applicable in the well-resolved sideband regime $\Omega_m\gg\kappa$, where $\kappa$ is the cavity linewidth.
In the opposite regime of a fast cavity $\kappa\gg\Omega_m$ one must resort to stroboscopic QND measurements, requiring interaction times $\ll\Omega_m^{-1}$~\cite{caves1980,braginsky1980}.

To date, such two-tone BAE measurement have exclusively been demonstrated in microwave optomechanical systems~\cite{hertzberg2010,suh2014}, 
where they have also been utilized to perform tomography of states produced by schemes that produce reservoir-engineered squeezed~\cite{kronwald2013,lecocq2015,lei2016} and entangled~\cite{woolley2014,ockeloen-korppi2018} mechanical states.
Yet, in all these experiments noise resulting from the use of a microwave amplifiers at elevated temperatures, resulted in substantially decreased efficiency and hindered beating the SQL~\cite{suh2014}.
Additionally, thermal noise at microwave frequencies can be non-negligible even at cryogenic temperatures, and requires careful calibration~\cite{weinstein2014}.
In contrast, optical homodyne or heterodyne detection is quantum-limited, and light is effectively a zero-temperature bath, allowing self-calibrated measurements of motion~\cite{purdy2015,underwood2015,qiu2018}.
Optomechanical systems using laser light have demonstrated quantum effects up to room temperature~\cite{purdy2017,sudhir2017}.
To date however, despite advances in operating in the QBA dominated regime in cavity optomechanics, BAE measurements in the optical domain have not been reported.
BAE measurements are compounded by instabilities arising from the excitation of higher-order mechanical modes due to the $2\Omega_m$ intensity modulation, and are susceptible to parametric instabilities~\cite{suh2012,suh2013,kampel2017}.


Here, we demonstrate
a two-tone BAE measurement in the optical domain of an oscillator in a thermal state (average occupation $\bar n$), using quantum-limited balanced heterodyne detection (BHD).
We first consider theoretically the scenario depicted in Fig.~\ref{fig:psd}b--d, in which a cavity optomechanical system is interrogated 
with two tones detuned by $\pm(\Omega_m+\delta)$ from the cavity resonance
and the two sidebands are detected using BHD.
In the case of a quantum-limited laser, the well-resolved sideband regime $\Omega_m\gg\kappa$, and within the rotating-wave approximation (RWA), the measured PSD is given by~\cite{SM}
\begin{equation}
\begin{split}
\bar S_{II}(\omega) &= 1+\eta\Geff^2\C 
         			    \Bigl[
           				  \bar{n}|\chi_m(\omega-\delta)|^2
           				  +(\bar{n}+1)|\chi_m(\omega+\delta)|^2\\
           				  &\quad+\C|\chi_m(\omega-\delta)-\chi_m(\omega+\delta)|^2
         				\Bigr],
\end{split}
\label{eq:Shet}
\end{equation}
where $\chi_m(\omega)=(-i\omega+\Geff/2)^{-1}$ is the mechanical susceptibility of the oscillator with total mechanical linewidth $\Geff$, $\eta$ the overall detection efficiency, and $\C=4g_0^2n_p/\kappa\Geff$ the optomechanical cooperativity proportional to the input power.
Here, $g_0$ is the vacuum optomechanical coupling strength,
and $n_p$ the mean number of intracavity photons due to each probe.

The PSD in Eq.~\eqref{eq:Shet} is normalized to the vacuum noise level, given by the constant 1 in the first term.
The first and second terms in brackets correspond to the anti-Stokes and Stokes scattered motional sidebands, respectively, having the Lorentzian shape of the mechanical susceptibility.
These exhibit the well-known quantum sideband asymmetry~\cite{clerk2010,safavi-naeini2012,khalili2012,weinstein2014,purdy2015,underwood2015,borkje2016,qiu2018}, resulting from the ratio $(\bar n+1)/\bar n$ between absorption and emission rates.
The last term in brackets is the QBA due to quantum noise in the probe light.
When $\delta\gg\Geff$, QBA appears as heating of the oscillator, adding $\nba=\C$ mean quanta (Fig.~\ref{fig:psd}c).
The two QBA components, which result from interaction with the positive and negative frequency parts of the probing field, have opposite phase.
When $\delta=0$ QBA is cancelled, yielding a pristine measurement of the oscillator,
$\bar S_{II}(\omega)=1+\eta\Geff\C\bar S_{XX}(\omega)$
with
$\bar S_{XX}(\omega)= (\Geff/2)(2\bar n+1)|\chi_m(\omega)|^2$
(Fig.~\ref{fig:psd}d).
In this case $2\nba$ quanta are added to the complementary quadrature~\cite{clerk2008}
\footnote{Note that our definition of $\nba$ is half that of Ref.~\onlinecite{clerk2008}, in order to facilitate comparison with conventional position measurement using homodyne detection.}.
In principle, one can then increase signal-to-noise ratio, (i.e. measurement sensitivity) indefinitely, with no deleterious effects on the measurement, simply by increasing probing power~
\footnote{However, even barring other technical limitations, in Eq.~\eqref{eq:Shet} we have neglected bad-cavity effects,
$\bar n\s{bad}=(\kappa/4\Omega_m)^2\C$
where photons scattered out of resonance interact with counter-propagating terms (neglected in the RWA) to induce QBA~\cite{clerk2008}.
In our experiment $\C\lesssim 10$ and $\bar n\s{bad}\sim 10^{-2}$ is completely negligible.}.


We performed a BAE measurement in a silicon nano\-beam optomechanical crystal~\cite{chan2012}, shown in Fig.~\ref{fig:experimental_setup}a.
Optically, the device functions as a single-sided cavity with a partially transmitting input mirror.
Light is evanescently coupled from a tapered optical fiber into a waveguide that forms part of the nanobeam (coupling efficiency exceeds 50\%).
The optical resonance is at $1540\unit{nm}$ with a linewidth of $\kappa/2\pi=1.7\unit{GHz}$, of which $\kappa\s{ex}=0.3\kappa$ are extrinsic losses to the input mirror.
The optical mode is optomechanically coupled to a mechanical breathing mode of frequency $\Omega_m/2\pi=5.3\unit{GHz}$, strongly confined due to a phononic bandgap, and an intrinsic linewidth of $\Gamma\s{int}/2\pi=84\unit{kHz}$.
This places the system in the resolved sideband regime~\cite{schliesser2008}.
The optomechanical coupling parameter is $g_0/2\pi=780\unit{kHz}$.
Full details of the device design, the setup and the system are given in Ref.~\onlinecite{qiu2018}.

\begin{figure}
\includegraphics[scale=1]{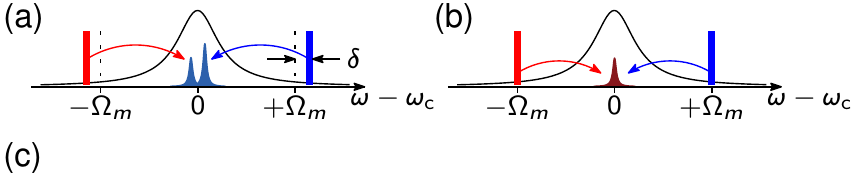}
\includegraphics[scale=1]{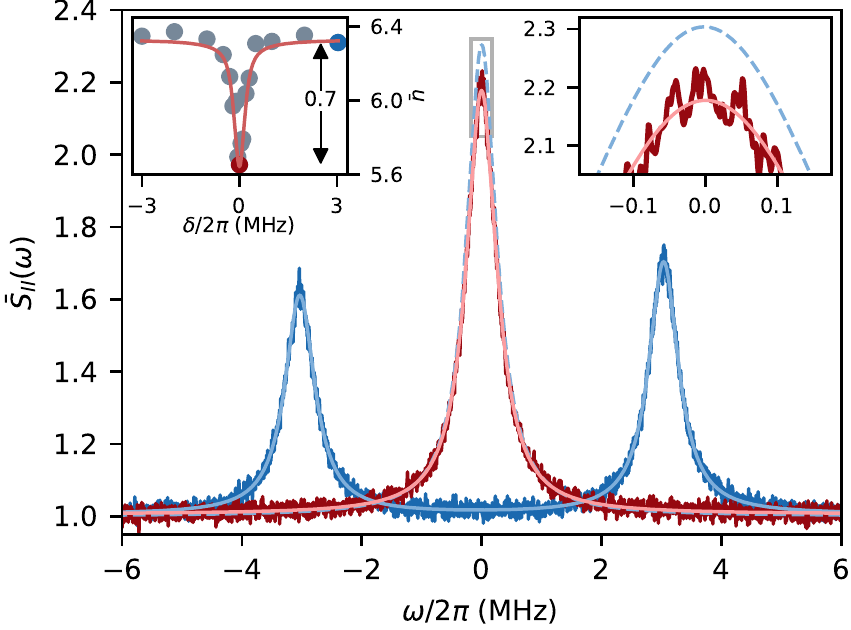}
\caption{\textbf{Experimental observation of backaction-evasion.}
(a)~and (b)~show non-BAE and BAE measurements, respectively, as explained in the main text.
(c)~Data traces, normalized to the vacuum noise level, for non-BAE (blue) and BAE measurements (red) are shown with Lorentzian fits.
The non-BAE sidebands exhibit motional asymmetry, used to self-calibrate the measurement in units of mechanical quanta.
The sum of the non-BAE sidebands, indicated in dashed blue, is larger than in the BAE case (red) by 0.7 mechanical quanta.
The right inset is a zoom of the indicated region.
The left inset shows the inferred occupation $\bar n$ as a function of the detuning $\delta$, with an analytic curve based on Eq.~\eqref{eq:Shet} with no free parameters.
In this measurement $n_p=290$, $n_c=320$, and $\C\s{cool}=3.8$.
}
\label{fig:bae}
\end{figure}
 
The system is placed in a $^3$He buffer gas cryostat
(Oxford Instruments HelioxTL),
allowing cryogenic operation down to $0.5\unit{K}$.
As detailed in Ref.~\onlinecite{qiu2018}, the buffer gas environment allows us to overcome the prohibitive optical absorption heating \textit{in vacuo} that has limited operation with these devices to very low photon numbers~\cite{meenehan2014} or pulsed operation~\cite{meenehan2015,riedinger2016,hong2017,riedinger2018}.
We are thus able to operate at high probe powers where QBA is observable.
The buffer gas causes additional damping, increasing the mechanical linewidth to
$\Gamma_m=\Gamma\s{int}+\Gamma\s{gas}$.
In addition to the BAE probes, we also apply a cooling tone red-detuned from the optical
resonance, to lower the thermal occupation $\bar n$ through optomechanical sideband cooling~\cite{schliesser2008},
$\bar n\simeq \nth/(1+\C\s{cool})$ with $\nth$ the occupation of the thermal environment.
The cooling tone also provides additional damping due to dynamical backaction,
$\Geff=\Gamma_m(1+\C\s{cool})$, which is the effective linewidth seen by the BAE probes.
Note that the balanced probes do not produce dynamical backaction.
Here, $\C\s{cool}=\C_0 n_c$ is the cooling tone cooperativity defined similarly to $\C$ but relative to the original linewidth $\Gamma_m$, with the single photon cooperativity $\C_0\equiv 4g_0^2/\kappa\Gamma_m$, and $n_c$ the mean intracavity photons due to the cooling tone.
The cooling tone is tuned $2\pi\times 220\unit{MHz}$ away from the red-detuned BAE probe to mitigate recently reported Kerr-type effects~\cite{qiu2018}.

The experimental setup is shown in Fig.~\ref{fig:experimental_setup}b.
The two BAE probes (as well as the cooling tone) are derived from two phase-locked lasers~\cite{SM}.
The three tones are combined in a free-space setup and coupled with the same polarization into a single mode fiber.
By blocking each beam path we ascertain equal power for each probe, stable to within 1\%.
The light reflected from the oscillator is directed to a BHD setup, where it is mixed with a local oscillator generated by a third laser.
As detailed in Ref.~\onlinecite{qiu2018}, by carefully characterizing our lasers we have determined that classical laser noise is negligible in our system.
Specifically we operate far from the relaxation-oscillation peak of our diode laser~\cite{kippenberg2013}.
Thus our detection is quantum limited, as in Eq.~\eqref{eq:Shet}.
In order to accurately tune the probes across the optical resonance, we temporarily 
switch the reflected light to a coherent response measurement setup~\cite{SM}.

Figure~\ref{fig:bae} shows BAE measurement of the mechanical oscillator, taken at a cryostat temperature of $2.0\unit{K}$ ($\nth\sim 7.9$) and buffer-gas pressure of $46\unit{mbar}$.
In this experiment, we vary the detuning $\delta/2\pi$ from $+3$ to $-3\unit{MHz}$.
The total mechanical linewidth across the measurement is $\Geff/2\pi=607\pm 7\unit{kHz}$ and the other measurement parameters are $n_p=290$, $n_c=320$, and $\C\s{cool}=3.8$.
When the probes are tuned away from the mechanical sidebands, $\delta/2\pi=3\unit{MHz}$,
the PSD exhibits motional sideband asymmetry that can be used to self-calibrate the measurement in terms of mechanical quanta (including QBA heating; see inset of Fig.~\ref{fig:bae}), $\bar n+\nba=6.3$ in this case~\cite{purdy2015,underwood2015,borkje2016,qiu2018}.
When tuning the probes on the mechanical sidebands, $\delta/2\pi=0\unit{MHz}$, the total thermomechanical noise is reduced by 0.7 mechanical quanta, in perfect agreement with independently calculated $\C=0.7$.
Thus more than 11\% of the noise in the non-BAE case is due to QBA.
This constitutes the first BAE measurement in the optical domain and the first with quantum-limited detection.

We now turn to discuss technical limitations of BAE measurement imposed by our system.
In conventional cavity-based position measurement employing homodyne detection~\cite{wilson2015,rossi2018}, one refers the on-resonance readout to mechanical quanta, 
i.e., expressing the peak of the measured PSD as
$\bar S_{II}^\mathrm{hom}(\Omega_m)\propto\nimp^\mathrm{hom} + \nba + (\bar n + \frac{1}{2})$,
where $\nimp^\mathrm{hom} = (16\eta\C)^{-1}$ is the measurement imprecision due to shot noise  (cf. Fig.~\ref{fig:psd}d).
The Heisenberg uncertainty relation requires
$4\sqrt{\nimp^\mathrm{hom}\nba}\geq 1$.
The SQL is achieved by minimizing the total added noise $\nadd=\nimp^\mathrm{hom}+\nba$ subject to this constraint, yielding $\nadd^\mathrm{SQL}=(4\eta)^{-1/2}$ ($\frac{1}{2}$ for ideal measurement).
In a BAE measurement there is no QBA component, however any device suffers extraneous heating due to optical absorption, adding excess heating backaction $\nba^\mathrm{th} = \beta\C$ analogous to $\nba$.
Additionally, in heterodyne detection $\nimp=(8\eta\C)^{-1}$, due to twice the vacuum noise compared to homodyne detection (`image band').
The minimum added noise is $\nadd^\mathrm{th} = \sqrt{\beta/2\eta}$.
When $\beta < \frac{1}{2}$, BAE outperforms conventional measurement of the same efficiency.

\begin{figure}[tbh]
\includegraphics[scale=1]{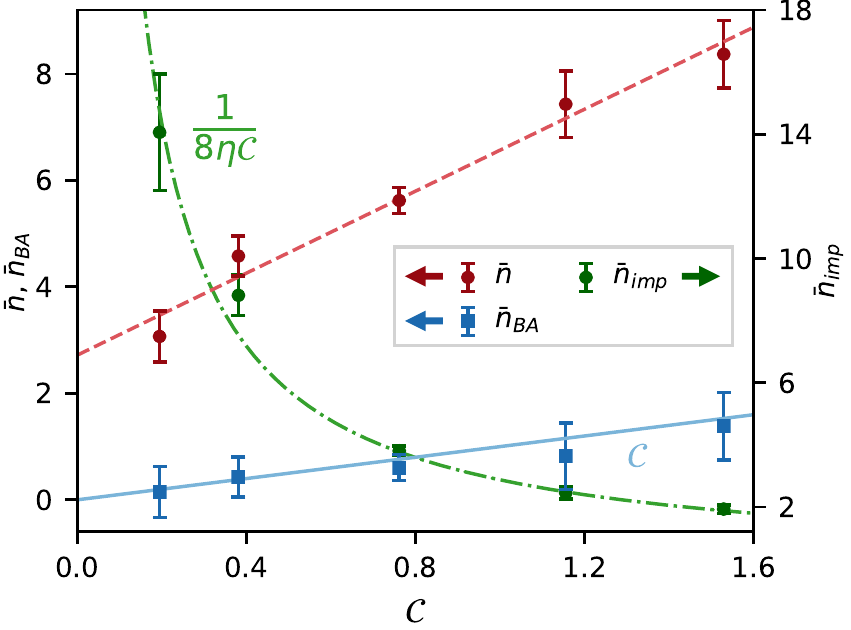}
\caption{\textbf{Effect of probe power on quantum backaction and optical absorption heating.}
The measurements were carried out at $1.6\unit{K}$ and $^3$He buffer-gas pressure of $30\unit{mbar}$,
with $n_c\simeq 420$ ($\C\s{cool}\simeq 5.0$).
The occupation $\bar n$ and the number of evaded QBA phonons $\nba$ vs.~independently-measured $\C$ are plotted on the left axis.
The error bars are due to uncertainty in occupancy calibrated using quantum sideband asymmetry.
The solid blue line plots $\nba=\C$.
The dashed red line is a linear fit to $\bar n$ with slope $\beta\C$ where $\beta=3.85$.
The right axis shows the imprecision noise with a fit $\nimp=1/8\eta\C$ yielding $\eta=0.04$.
}
\label{fig:power_sweep}
\end{figure}

Figure~\ref{fig:power_sweep} shows a set of measurements done at $1.6\unit{K}$ ($\nth\simeq 6.3$) and buffer-gas pressure of $30\unit{mbar}$ with variable probe power $n_p$ and constant cooling tone power $n_c=420$ (set by the maximum probe power).
Both $\bar n$ and $\nba$ are plotted against the independently measured cooperativity $\C$, with $\nba$ in excellent agreement with theory (blue solid line, slope of 1), with maximum cancellation of $\nba=1.4$ out of $\bar n+\nba=9.8$ quanta, or 14\% (reduction of $0.67\unit{dB}$).
The linear fit to $\bar n$ yields $\beta = 3.85$.
Thus, although QBA \emph{is evaded} in our measurement, extraneous heating is still a limiting factor, as can be seen directly from Fig.~\ref{fig:power_sweep}.
The imprecision noise is also in excellent agreement with theory and yields $\eta=0.04$, in agreement with previous measurements~\cite{qiu2018} of the same system.
Thus $4\sqrt{\nimp\nba^\mathrm{th}}=\sqrt{2\beta/\eta} = 13.88$.
Compared to a measurement at the SQL with the same efficiency, the optimal added noise is $\nadd^\mathrm{th}=2.78\times\nadd^\mathrm{SQL}$.

In conclusion, we have explicitly demonstrated evasion of QBA for the first time in the optical domain, an important step for various quantum measurements with nanomechanical oscillators in the sideband resolved regime.
Though the current generation of devices is limited by low efficiency and extraneous heating,
improvements in design and fabrication already yield an intrinsic optical $Q$-factor improvement by a factor of $\sim 5$, addressing both deficiencies~\cite{Qiu2018b}.
This opens the path for creating motional squeezed states through reservoir engineering~\cite{kronwald2013} demonstrated so far only in the microwave domain~\cite{lecocq2015,wollman2015} and generation of squeezed light through mechanical dissipation~\cite{kronwald2014} which remains elusive.


\begin{acknowledgments}
  \paragraph{Acknowledgments.} IS acknowledges support by the European Union's Horizon 2020 research and innovation programme under Marie Sko\l{}odowka-Curie IF grant agreement No.~709147 (GeNoSOS).
  LQ acknowledges support by Swiss National Science Foundation under grant No.~163387.
  DM acknowledges support by the UK Engineering and Physical Sciences Research Council (EPSRC) under Grant No.~EP/M506485/1.
  AN acknowledges a University Research Fellowship from the Royal Society and support from the Winton Programme for the Physics of Sustainability.
  TJK acknowledges financial support from an ERC AdG (QuREM).
  This work was supported by the SNF, the NCCR Quantum Science and Technology (QSIT), and the European Union's Horizon 2020 research and innovation programme under grant agreement No.~732894 (FET Proactive HOT).
  All samples were fabricated in the Center of MicroNanoTechnology (CMi) at EPFL.
\end{acknowledgments}

\paragraph{Data Availability.} All data and analysis files will be made available via
\texttt{zenodo.org} upon publication.

\bibliography{refs}



\makeatletter
\close@column@grid
\clearpage

\onecolumngrid
\begin{center}
  \textbf{\large 
    \@title\\[0.5cm]
  	Supplemental Material\\[.5cm]
  }
  Itay Shomroni,$^{1,*}$ Liu Qiu,$^{1,*}$ Daniel Malz,$^2$ Andreas Nunnenkamp,$^2$ and Tobias J. Kippenberg$^{1,\dagger}$\\[.1cm]
  {\itshape 
  	$^1$Institute of Physics, \'Ecole Polytechnique F\'ed\'erale de Lausanne, Lausanne 1015, Switzerland\\
	$^2$Cavendish Laboratory, University of Cambridge, Cambridge CB3 0HE, United Kingdom
  }\\
  $^\dagger$Electronic address: tobias.kippenberg@epfl.ch\\
(Dated: \@date)\\[1cm]
\end{center}
\twocolumngrid

\makeatother

\setcounter{equation}{0}
\setcounter{figure}{0}
\setcounter{table}{0}
\setcounter{page}{1}
\renewcommand{\theequation}{S\arabic{equation}}
\renewcommand{\thefigure}{S\arabic{figure}}
\renewcommand{\bibnumfmt}[1]{[S#1]}
\renewcommand{\citenumfont}[1]{S#1}

\section{Theory}

The theory of dual-tone backaction-evading measurements in optomechanics is already well established~\cite{clerk2008}, but we repeat the key elements here for convenience of the reader.
The system is described by the Hamiltonian
\begin{multline}
\hat H = \hbar\omega_c\hat a\dagg\hat a 
       + \hbar\Omega_m\hat b\dagg\hat b
       - \hbar g_0\hat a\dagg\hat a(\hat b\dagg + \hat b) 
       + \hat H\s{drive}
\label{eq:Hamiltonian}
\end{multline}
where $\hat a$ and $\hat b$ the annihilation operators of a cavity photon and a mechanical phonon, respectively.
The cavity is driven by a coherent drive
$\alpha\s{in}(t) = (\alpha_{+}e^{-i\Omega t} + \alpha_{-}e^{i\Omega t})e^{-i\omega_l t}$
with carrier frequency $\omega_l = \omega_c+\Delta$ and amplitude-modulated at frequency $\Omega=\Omega_m+\delta$, giving
$\hat H\s{drive} = i\sqrt{\kappa}[\alpha\s{in}(t)\hat a\dagg - \alpha\s{in}^*(t)\hat a]$.

We follow standard procedure in cavity optomechanics~\cite{aspelmeyer2014}.
We move to the interaction picture with respect to the Hamiltonian
$\hat H_0 = \hbar\omega_l\hat a\dagg\hat a + \hbar\Omega\hat b\dagg\hat b$
and linearize the operators $\hat a\rightarrow \bar a+\da$, $\hat b\rightarrow\bar b+\db$.
In this rotating frame we can write
$\hat H = \hat H\s{RWA} + \hat H\s{CR}$
with
\begin{multline}
\hat H\s{RWA}/\hbar = -\Delta\da\dagg\da - \delta\cdot \db\dagg\db \\
 - \bigl[ (g_{+}\db\dagg + g_{-}\db)\da\dagg + (g_{+}\db + g_{-}\db\dagg)\da \bigr]
\end{multline}
with $g_\pm=g_0\bar a_\pm$ the drive-enhanced coupling, where $\bar a_\pm$ is the intracavity amplitude due to each drive tone.
The counter-rotating  Hamiltonian
\begin{multline}
\hat H\s{CR}/\hbar = -\bigl[g_{+}e^{-2i\Omega t}\db + g_{-}e^{2i\Omega t}\db\dagg\bigr]\da\dagg \\
               -\bigl[g_{+}e^{2i\Omega t}\db\dagg + g_{-}e^{-2i\Omega t}\db\bigr]\da
\end{multline}
contains off-resonant terms and can be neglected in the sideband-resolved regime $\Omega_m\gg\kappa$. Exact analytical solution is possible in the general case~\cite{malz2016b}.
Bad-cavity effects in BAE measurements were considered in Ref.~\onlinecite{clerk2008}.
Including the coupling to the mechanical and optical baths and using standard input-output theory leads to the quantum Langevin equations~\cite{gardiner2004}
\begin{subequations}
\label{eq:Langevin}
\begin{align}
\label{eq:LangevinOptical}
\delta\dot{\hat a} &= -(\kappa/2-i\Delta)\da + i(g_{-}\db+g_{+}\db\dagg) + \sqrt{\kappa}\da\s{in} \\
\delta\dot{\hat b} &= -(\Geff/2-i\delta)\db  + i(g_{-}\da+g_{+}\da\dagg) + \sqrt{\Geff}\db\s{in},
\end{align}
\end{subequations}
where $\Geff$ is the dissipation rate of the mechanical oscillator and we have introduced the optical ($\da\s{in}$) and mechanical ($\db\s{in}$) input noise operators.
We have assumed for simplicity no intrinsic optical losses (highly overcoupled cavity).
Note that in the rotating frame, the mechanical quadrature operators are given by 
$\hat X = \frac{1}{\sqrt{2}}(e^{i\delta t}\db\dagg + e^{-i\delta t}\db)$
and
$\hat Y = \frac{i}{\sqrt{2}}(e^{i\delta t}\db\dagg - e^{-i\delta t}\db)$.
When $g_{+}=g_{-}$ and $\delta=0$, the optical field couples exclusively $\hat X$ [Eq.~\eqref{eq:LangevinOptical}], the key feature of BAE measurement.

By transforming the Langevin Eqs.~\eqref{eq:Langevin} to Fourier space, we can relate the field operators in simple matrix form
$\mathbf{d}(\omega) = \pmb{\chi}(\omega)\mathbf{L\,d\s{in}}(\omega)$
with
$\mathbf{d} = (\da,\,\da\dagg,\,\db,\,\db\dagg)^T$,
$\mathbf{d\s{in}} = (\da\s{in},\,\da\s{in}\dagg,\,\db\s{in},\,\db\s{in}\dagg)^T$,
$\mathbf{L} = \mathrm{diag}(\sqrt{\kappa},\,\sqrt{\kappa},\,\sqrt{\Geff},\,\sqrt{\Geff})$, and
\begin{widetext}
\begin{equation}
\pmb\chi(\omega) =
\begin{pmatrix}
\chi_c^{-1}(\omega+\Delta) & 0                          & -ig_{-} & -ig_{+} \\
0                          & \chi_c^{-1}(\omega-\Delta) &  ig_{+} &  ig_{-} \\
-ig_{-}                    & -ig_{+}             & \chi_m^{-1}(\omega+\delta) & 0 \\
 ig_{+}                    &  ig_{-}             & 0 & \chi_m^{-1}(\omega-\delta)
\end{pmatrix}^{-1}
\quad \text{with} \quad
\left\{
\begin{aligned}
\chi_c(\omega) &= \frac{1}{-i\omega+\kappa/2} \\
\chi_m(\omega) &= \frac{1}{-i\omega+\Geff/2}
\end{aligned}
\right.
\end{equation}
\end{widetext}
The output fields are given by the input-output relations,
e.g.,~$\da\s{out} = \da\s{in} - \sqrt{\kappa}\da$, yielding the matrix equation
$\mathbf{d\s{out}} = [1 - \mathbf{L\,\pmb\chi(\omega)\,L}]\mathbf{d\s{in}}$,
with 
$\mathbf{d\s{out}} = (\da\s{out},\,\da\s{out}\dagg,\,\db\s{out},\,\db\s{out}\dagg)^T$.

The output optical field is detected using balanced heterodyne detection, mixing it with a strong local oscillator with frequency $\omega_l+\Delta\s{LO}$ (in the lab frame) on a beam\-splitter and subtracting the detected intensity from the two beam\-splitter output arms.
This yields photocurrent with symmetrized PSD~\cite{bowen2016}
\begin{equation}
\bar S_{II}(\omega) \propto S_{\da\s{out}\da\s{out}}(\Delta\s{LO}+\omega)
                       + S_{\da\s{out}\dagg\da\s{out}\dagg}(\Delta\s{LO}-\omega).
\label{eq:heterodyneBowen}
\end{equation}
Using the solutions for $\da\s{out}$, $\da\s{out}\dagg$ and the correlations of the input noise operators
\begin{subequations}
\label{eq:inputCorrelations}
\begin{align}
\langle\da\s{in}\dagg(\omega)\da\s{in}(\omega')\rangle & = 0 \\
\langle\da\s{in}(\omega)\da\s{in}\dagg(\omega')\rangle & = \delta(\omega+\omega')
\label{eq:shotNoise} \\
\langle\db\s{in}\dagg(\omega)\db\s{in}(\omega')\rangle & = \bar n\delta(\omega+\omega') \\
\langle\db\s{in}(\omega)\db\s{in}\dagg(\omega')\rangle & = (\bar n+1)\delta(\omega+\omega')
\end{align}
\end{subequations}
with $\bar n$ the mean thermal occupation of the environment seen by the oscillator, 
the PSD~\eqref{eq:heterodyneBowen} can be evaluated.
Apart from a white noise floor due to shot noise~\eqref{eq:shotNoise}, both 
$S_{\da\s{out}\da\s{out}}(\omega)$ and $S_{\da\s{out}\dagg\da\s{out}\dagg}(\omega)$
contain information near $\omega\approx\Geff,\delta\ll\Delta\s{LO}$.
Hence the PSD~\eqref{eq:heterodyneBowen} will contain information near $\Delta\s{LO}$ dominated by one of them.
Henceforth we refer all measured PSDs to $\Delta\s{LO}$ (i.e. Fig.~1c,d in the main text).

We now specialize to the case $\Delta=0$ and $g_{+}=g_{-}$, as in our experiment.
We can also approximate $\chi_c(\omega)\approx\chi_c(0)$ since for our frequencies of interest $\omega\ll\kappa$.
Including finite detection efficiency $\eta$ finally yields the PSD given in the main text, normalized to the vacuum noise level,
\begin{equation}
\begin{split}
\bar S_{II}(\omega) &= 1+\eta\Geff^2\C 
         			    \Bigl[
           				  \bar{n}|\chi_m(\omega-\delta)|^2
           				  +(\bar{n}+1)|\chi_m(\omega+\delta)|^2\\
           				  &\quad+\C|\chi_m(\omega-\delta)-\chi_m(\omega+\delta)|^2
         				\Bigr]
\end{split}
\end{equation}
with optomechanical cooperativity $\C=4g_0^2n_p/\kappa\Geff$ and $n_p = \lvert\bar a\rvert^2$.
For $\delta=0$, the quadrature $\hat X$ is given by
$\hat X= \sqrt{\Geff/2}\,\chi_m(\omega)[b\s{in}\dagg(\omega) + b\s{in}(\omega)]$ with PSD $\bar S_{XX}(\omega)= (\Geff/2)(2\bar n+1)|\chi_m(\omega)|^2$.

\section{Frequency Setup and Phase Lock}

\begin{figure}[tbh]
\includegraphics[scale=1]{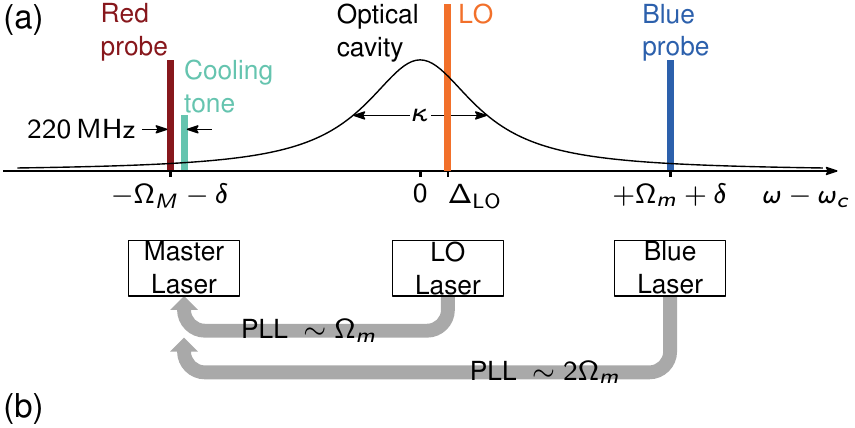}
\includegraphics[scale=1]{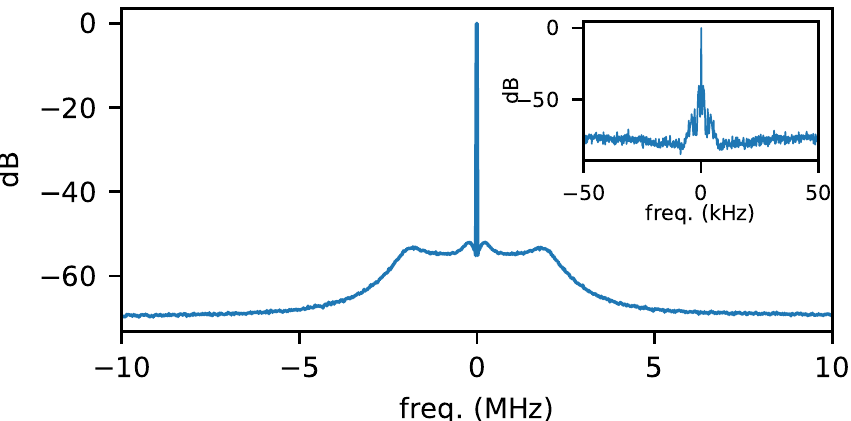}
\caption{
	\textbf{Complete frequency setup and phase-locked loop beat note.}
	(a)~The relation of the various tones used in the experiment to the cavity resonance.
	The LO does not enter the cavity.
	(b)~A typical out-of-loop PLL beat note, relative to the offset frequency.
	The resolution bandwidth (RBW) is~$6.25\unit{kHz}$.
	The inset shows a zoom-in with RBW of~$31.25\unit{Hz}$.
}
\label{fig:pll}
\end{figure}

Figure~\ref{fig:pll}a shows the complete picture of the various laser tones applied in the experiment.
A master laser generates the red-detuned probe and, via acousto-optic frequency shifter, the cooling tone.
Two other lasers are referenced to the master laser through a phased-locked loop (PLL).
One laser generates the local oscillator (LO) for the balanced heterodyne detection and is locked at an offset of $\sim\Omega_m$ to the blue of the master laser.
A second laser generates the blue-detuned probe and is locked at an offset $\sim 2\Omega_m$ to the blue of the master laser.

We perform the phase-lock using a PID controller with $10\unit{MHz}$ bandwidth to control the current of the diode lasers.
A typical beat note is shown in Fig.~\ref{fig:pll}b.
The residual phase error~\cite{zhu1993,prevedelli1995,marino2008}, computed from the ratio coherent to total power, is $\langle\sigma_\phi^2\rangle\simeq 5\times 10^{-3}\unit{rad^2}$, limited by the resolution bandwidth.

\section{Optomechanical Cooling and Absorption Heating}

At low temperatures, intracavity photons shift the optical resonance to higher frequencies, due to a combination of thermo-optic and thermal expansion effects in silicon.
The cavity is optically unstable when driven with the BAE probes alone (due to the blue-detuned probe), and an additional red-detuned (cooling) tone of sufficient power is required.
In our system we have found empirically that we need $n_c\gtrsim n_p/2$ for stable operation.

\section{Probe Tuning via Coherent Response}

\begin{figure}[tbh]
\includegraphics[scale=1]{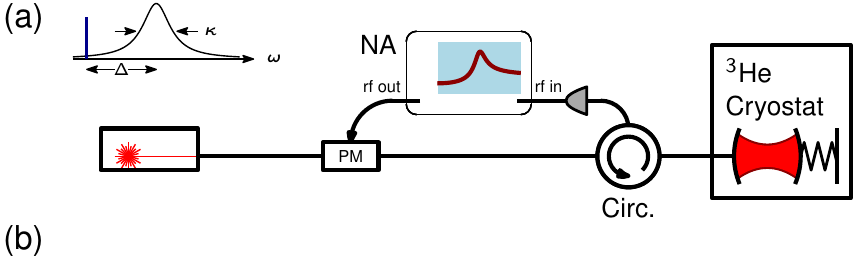}
\includegraphics[scale=1]{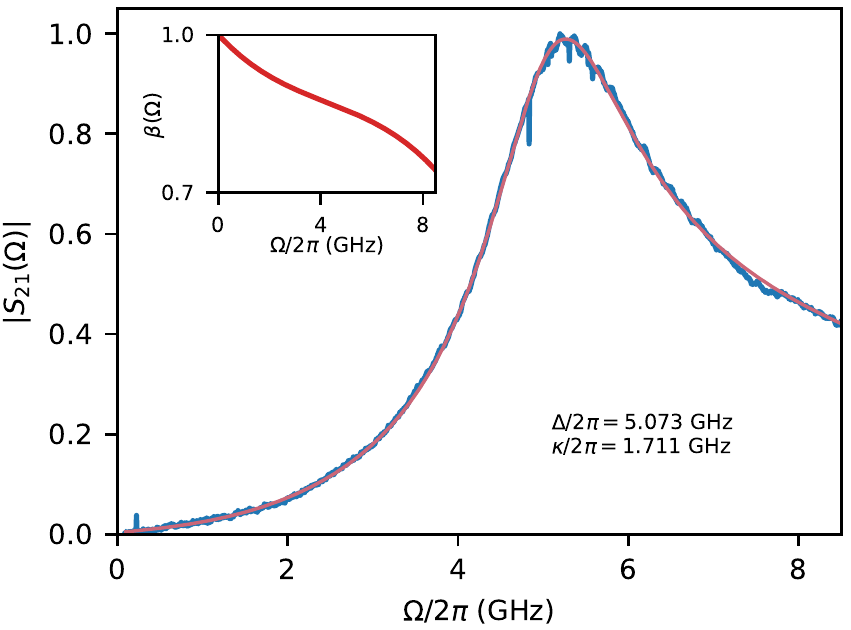}
\caption{
    \textbf{Coherent response determination of detuning and linewidth.}
    (a)~Simplified setup for optical measurements using coherent response (part of the full experimental setup).
    (b)~An example of a measurement with a fit Eq.~\eqref{eq:s21}, yielding $\kappa$ and $\Delta$. A correction due to extraneous frequency-dependent response, measured and shown in the inset, is included in the fit.	
}
\label{fig:na}
\end{figure}

In experiments such as this, it is of utmost importance to tune the probes accurately around the optical resonance.
Active locking to the cavity (e.g. using a Pound-Drever-Hall technique) are inappropriate for a single-sided cavity, and also result in driving the cavity on resonance.
We use passive tuning of our master laser, from which the other tones are derived, using the cavity coherent response, similar to previous experiments~\cite{verhagen2012}.
From the response curve we extract accurate values of both $\Delta$ and $\kappa$.
Here we give details on the method.
A simplified setup is shown in Fig.~\ref{fig:na}a.
The laser is phase-modulated using RF output of a network analyzer (NA).
The carrier and sidebands reflected from the cavity interfere on a fast photodetector and the photocurrent fed to the NA input, measuring the magnitude of the $S_{21}$ parameter.
The amplitude incident on the cavity is given by
\begin{equation}
a\s{in}(t) \simeq a_0\biggl(1 + \frac{\beta}{2}e^{i\Omega t} - \frac{\beta}{2}e^{-i\Omega t}\biggr),
\end{equation}
with $\beta$ the modulation index, and the reflected light is
\begin{equation}
a\s{out}(t) \simeq a_0\biggl[r(\Delta) + \frac{\beta}{2}r(\Delta+\Omega)e^{i\Omega t} - \frac{\beta}{2}r(\Delta-\Omega)e^{-i\Omega t}\biggr],
\end{equation}
with
\begin{equation}
r(\Delta)= 1 - \frac{\eta_c\kappa}{\kappa/2-i\Delta}
\end{equation}
the amplitude reflection coefficient at detuning $\Delta$ and $\eta_c\equiv\kappa\s{ex}/\kappa$ the cavity coupling parameter.
The magnitude of the $S_{21}$ parameter, the $\Omega$ frequency component of the photocurrent $|a\s{out}(t)|^2$, is given by (here and below we omit a constant scale factor)
\begin{equation}
|S_{21}(\Omega)| = \frac{\beta}{2}|r(\Delta)r^{*}(\Delta-\Omega)-r^{*}(\Delta)r(\Delta+\Omega)|.
\label{eq:s21}
\end{equation}

Figure~\ref{fig:na}b shows a typical coherent response measurement, which deviates significantly from a Lorentzian when $\Delta\sim\kappa$, as in our case.
Additionally, when scanning over a wide bandwidth, one has to take into account the frequency dependence of $\beta$ (due to phase modulator, rf cables, detector response etc.).
A robust and reliable procedure to calibrate the frequency dependence of the entire detection chain is to take several traces at various detunings, and fit all of them simultaneously to Eq.~\eqref{eq:s21} with only $\Delta$ variable across traces, and with a high-order polynomial in $\Omega$ as a pre\-factor.
This pre\-factor is then applied in all subsequent fits.
The inset in Fig.~\ref{fig:na}b shows the frequency dependence of $\beta$ given by the polynomial.
We adjust probe detuning using this method prior to each data point acquisition.
By repeatedly acquiring $\Delta$ in a single instance, we estimate our accuracy to be $\pm 20\unit{MHz}$ ($\pm 0.01\kappa$).


\end{document}